\documentstyle[12pt,aasms4]{article}
\def\ltsima{$\; \buildrel < \over \sim \;$}
\def\lsim{\lower.5ex\hbox{\ltsima}}
\def\gtsima{$\; \buildrel > \over \sim \;$}
\def\gsim{\lower.5ex\hbox{\gtsima}}

\begin{document}
\title{OPTICAL VARIABILITY IN ACTIVE GALACTIC NUCLEI:
STARBURSTS OR DISK INSTABILITIES?}

\author{T. Kawaguchi,\quad{\rm and}\quad S. Mineshige}

\affil{Department of Astronomy, Kyoto University, Sakyo-ku, 
 Kyoto 606-8502, Japan; 
 kawaguti@kusastro.kyoto-u.ac.jp, minesige@kusastro.kyoto-u.ac.jp}

\author{M. Umemura}

\affil{Center for Computational Physics, University of Tsukuba, 
 Tsukuba, Ibaraki 305-0006, Japan;
 umemura@rccp.tsukuba-u.ac.jp}

\and 

\author{Edwin L. Turner}

\affil{Princeton University Observatory, Peyton Hall, Princeton, NJ 08544; 
 elt@astro.princeton.edu}


\begin{abstract}

Aperiodic optical variability is a common property of
Active Galactic Nuclei (AGNs), though its physical origin 
is still open to question. 
To study the origin of the optical -- ultraviolet 
variability in AGN, we compare light curves of 
two models to observations of quasar 0957+561 in terms of a 
structure function analysis.
In the starburst (SB) model, random superposition of supernovae 
in the nuclear starburst region
produce aperiodic luminosity variations, 
while in the disk-instability (DI) model, variability is caused by 
instabilities in the accretion disk around a 
supermassive black hole.
We calculate fluctuating light curves and structure functions, $V(\tau)$, 
by simple Monte-Carlo simulations on the basis of the two models.
Each resultant $V(\tau)$ possesses a power-law portion, 
$[V(\tau)]^{1/2} \propto \tau^{\beta}$, at short time lags ($\tau$).
The two models can be distinguished by the logarithmic slope, $\beta$;
$\beta \sim$ 0.74--0.90 in the SB model and 
$\beta \sim$ 0.41--0.49 in the DI model, 
while the observed light curves exhibit $\beta \sim$ 0.35.
Therefore, we conclude that the DI model is favored over 
the SB model to explain the slopes of the observational 
structure function, in the case of 0957+561, though this object 
is a radio-loud object and thus not really a fair test for the SB model.
In addition, we examine the time-asymmetry of the light curves by 
calculating $V(\tau)$ separately for 
brightening and decaying phases.
The two models exhibit opposite trends of time-asymmetry to some extent,
although the present observation is not long enough 
to test this prediction.

\end{abstract}


\section{INTRODUCTION}

Emission from Active Galactic Nuclei (AGNs) has long been
known to exhibit rapid and apparently random variability,
over wide wavelength ranges from radio to X ray or $\gamma$ ray
(Krolik et al. 1991; Edelson et al. 1996).  The fluctuation
power spectrum has a power-law dependence on frequency,
$\propto f^{-\alpha}$ with $\alpha$ = 1.0 $\sim$ 2.5
(Lawrence \& Papadakis 1993; Green, McHardy \& Lehto 1993; 
Leighly \& O'Brien 1997; Hayashida et al. 1997).
Simultaneous multi-wavelength observations have also revealed 
that there is strong time correlation between
optical -- ultraviolet (UV) and X-ray emission with little delay
(Clavel et al. 1992; Edelson et al. 1996; Warwick et al. 1996),
while the variation of the H{$\beta$} emission line follows
the optical continuum variation with a delay of $\sim$ 20 days 
(Peterson et al. 1994; Netzer \& Peterson 1997 
for emission line variability).  
At present, many 
multi-wavelength monitoring projects are in progress
(e.g. Kundi\'{c} et al. 1997; O'Brien \& Leighly 1997 and 
references therein).
In spite of these successful observations, 
the location of regions emitting in various energy bands
and physical origin of the variability are still open to question.
In addition, the apparent time-reversibility of AGN light curves 
has been recognized as a potential problem recently 
(Aretxaga 1997).
Therefore, one of the major goals of variability studies is
to identify and characterize the physical processes responsible 
for the observed variability.

The presence of huge energy outputs and high-energy emission 
strongly indicates the existence of a supermassive black hole
and a surrounding accretion disk in AGN (e.g. Rees 1984).
This picture is supported by a number of observations; e.g. 
the detection of large peculiar velocity dispersion of the stars 
near the nucleus, substantial, short-term X-ray variability 
and asymmetric broad Fe fluorescence line features (Tanaka et al. 1995). 
Moreover, it has been suggested that the optical -- ultraviolet emission 
may mainly come from the accretion disk, which is supported
indirectly by the rough agreement between the theoretical spectral 
energy distribution calculated by the standard 
accretion disk model and the observational ones (Shields 1978; Malkan 1983).
However, since the standard optically thick accretion disk 
is too cool to produce any X-ray emission, 
another mechanism is required.
Regarding the relation between these two energy sources, 
there is the following suggestion. 
The absence of time lag between UV and X-ray variability 
suggests reprocessing;
the UV emission comes from the optically thick disk which is partially 
illuminated by an X-ray source 
(i.e. accretion disk atmosphere) and is also heated by internal
viscosity and accretion (Santos-Lle\'{o} et al. 1995, 
Edelson et al. 1996).
Occasional flare events or blob formation caused by some 
instability in the atmosphere should produce a luminosity variation in time.
This is the basic idea behind
the disk-instability model (hereafter DI model).

There exist, however, completely different models for radio-quiet AGN 
which do not always involve black holes.
The most popular among them is the starburst model (hereafter SB model;
e.g., Terlevich et al. 1992). 
In this model the energy is generated by violent star 
formation activity in the innermost regions of AGN. 
The observed flux variability of AGNs is ascribed to transient
phenomena associated with the evolution of massive stars and 
supernovae (SNe).
There is no difficulty in accounting for the average bolometric 
luminosity of AGN in terms of SNe; for example a Seyfert galaxy
luminosity of $10^{44}$ erg s$^{-1}$ can be accommodated by $\sim$ 1 
SN per year (the energy released by one supernova being 
$\sim \ 10^{51}$ ergs), whereas a high luminosity QSO would require 
a few hundred SNe per year. 
However, it would appear to be difficult to explain the rapid 
X-ray variability with the SB model.


To answer these questions, 
we have analyzed the optical variability of quasar 0957+561 
and each of the SB and DI models, 
using a first-order structure function analysis.
The main goal of the present study is to clarify distinction between 
variability caused by starburst (SB) and disk instability (DI),
and consequently to derive some constraints on plausible models. 
Structure function analysis provides a method of quantifying 
time variability without the problems of windowing, aliasing, etc., 
that are encountered in the traditional Fourier analysis technique. 
It potentially provides good information on the nature of the process
that causes variation.
This technique has been used previously by a number of authors for 
the study of time series 
(e.g., Simonetti, Cordes \& Heeschen 1985; 
Hughes, Aller \& Aller 1992; Press, Rybicki \& Hewitt 1992).

The best available QSO optical light curve, in terms of sampling rate, data
volume and photometric accuracy, was obtained by Kundi\'{c} et al. (1995, 1997)
for the gravitational lens system 0957+561A,B.  The data were obtained for
the purpose of determining the lensing differential time delay between the
two images of the background quasar, a measurement which had generated
considerable controversy over the years (see Haarsma et al. 1997).  The
clear determination of the delay (Kundi\'{c} et al. 1997), which effectively
ended the controversy, makes the system even more suitable for our purposes
since the light curves of the two images can now be combined to produce
one effectively even longer and better sampled light curve of the source
itself.  Moreover, comparison of the light curves of the two images allows
variations intrinsic to the source (in which we are interested) to be
distinguished from those produced by any process occurring along the line
of sight (such as microlensing or extinction variations with time, in
which we are not interested).  In other words, variations intrinsic to the
source will occur in both images, though delayed in image B relative to
A, while those which are extrinsic will only be seen in one image.  This
technique indicates that the observed variations are entirely, or almost
entirely, due to intrinsic source variations and that any extrinsic
variations have photometric amplitudes no larger than a few hundredths of
a magnitude, comparable to the photometric noise.  Finally, the fact that
gravitational lensing occurs due to random chance alignments of foreground
and background objects implies that there is no bias due to analyzing a
lens system; the source quasar in 0957+561A,B is essentially a randomly
chosen (by nature) object.  Note, however, that 0957+561 is a radio
loud quasar and thus not an entirely fair test of the SB model.  
(In this context, we do not preclude the presence of a black hole 
in our SB model,
since otherwise radio emission of 0957+561 is cannot be accounted for.)
Nevertheless,
we confine the present analysis to this best available data set but
note that no firm conclusions can be achieved until good quality light
curves are available for a statistically valid sample of AGN.


The descriptions of models and structure functions including the numerical 
results are given in section 2.
In section 3 we describe the time-asymmetry in the 
observed and predicted light curves,
and section 4 is devoted to discussion.
Finally, we summarize our conclusions in section 5.

\section{LIGHT CURVES AND STRUCTURE FUNCTION}

\subsection{Structure function analysis}

Here we use only the first-order structure function described below.
The general definition of structure functions and some of their 
properties are given by Simonetti et al. (1985). 
For a time series of optical magnitude 
[ $m_{\rm opt}(t_i), \ i = 1, 2, \ldots ; \ t_i \, < \, t_j $ ], 
the first-order structure function, $V(\tau)$, 
and the autocorrelation function, $C(\tau)$, 
are defined as
\begin{eqnarray}
 V(\tau) \equiv \frac{1}{N(\tau)}
  \sum_{i<j}\left[\, m_{\rm opt}(t_i) \, - \, m_{\rm opt}(t_j) \,\right]^2
\end{eqnarray}
\begin{eqnarray}
 C(\tau) \equiv \frac{1}{N(\tau)}
  \sum_{i<j}\, m_{\rm opt}(t_i) \times m_{\rm opt}(t_j)
\end{eqnarray}
Summation is made over all pairs $(i,j)$ for which $t_j - t_i = \tau$, 
and $N(\tau)$ denotes the number of such pairs.
For a stationary random process the structure function is related to 
its autocorrelation function by
\begin{eqnarray}
 V(\tau) = 2 \, [\,C(0) - C(\tau)\,].
\end{eqnarray}

The typical shape of the measured structure functions basically 
consists of three distinct parts (cf. Hughes et al. 1992).
First, there is a plateau,
at time lags longer than the longest correlation timescale, 
with a value of twice the variance of the fluctuations.
Second, there is another plateau, at short time lags, with a value 
corresponding to twice the variance of the measurement noise
(which is absent in model calculations).
Finally, the intermediate part between the two plateaus exhibits
a power-law increase with increasing time lag;
\begin{eqnarray}
[\,V(\tau)\,]^{\frac{1}{2}} \propto \tau^{\beta}.  
\end{eqnarray}

In fact, the structure functions for the optical light curves 
of many AGNs increase steadily for $\tau \lsim$ 2--3 yr
and flatten at larger lags (Hook et al. 1994; Tr\`{e}vese et al. 1994; 
Cristiani et al. 1996; Cid Fernandes et al. 1997). 
Its logarithmic slope ($\beta$) depends on the nature of the intrinsic 
variation of the source (e.g., shot noise, flicker noise, and so on).
Kundi\'{c} et al. (1997) report the results of photometric monitoring 
of 0957+561 A and B during about two years.
The light curve and structure function of their data set 
are displayed in Figure 1, where $[V(\tau)]^{1/2}$ is calculated 
by averaging the structure functions of image A and B for 
each time lag $\tau$.
In Figure 1, $[V(\tau)]^{1/2}$ shows a logarithmic slope of 
$\beta \ \sim$ 0.35 at 2 $\lsim \tau \lsim$ 500 day.
Independently, Press, Rybicki \& Hewitt (1992) previously 
calculated the structure function for the same object 
based on the data available at that time
and derived the logarithmic slope to be $\beta \sim$ 0.27--0.34.

Although we do not know the logarithmic slopes ($\beta$) 
of structure functions for optical light curves of a representative 
sample of AGN, the range of slopes $\beta$ can be estimated by the slopes of 
power density spectra (PDS),
using the relation between the structure function and PDS as follows:
The typical PDS [\,$P(f)$\,] is also described 
by a  power-law, $P(f) \propto f^{- \alpha} $, and the power-law
index ($\alpha$) is related to that of the structure function by 
$\alpha = 1 + 2 \beta \, $ for $1 \leq \alpha \leq 3$ 
again under the condition that the time 
series is stationary (Hughes et al. 1992). 
For example, we have $\beta \sim 0.25-0.40$
for $\alpha \sim 1.5-1.8$ (Lawrence \& Papadakis 1993; 
Hayashida et al. 1997 for X-ray variability data).
If a given time series is produced by multiple superposition of 
some canonical shot (with a fixed time profile) at random in time, 
the analyzed value of $\beta$ and $\alpha$ are determined solely by 
the profile of the canonical shot (see also Aretxaga, 
Cid Fernandes \& Terlevich 1997; Press \& Rybicki 1997).
If the frequency of shots depend on their magnitudes
(e.g., small shots are frequent, whereas large shots are rare),
the resultant PDS and structure function are influenced 
by the distribution of the size (peak intensity and duration)
of individual shots
(see Press 1978; Takeuchi, Mineshige \& Negoro 1995).

Incidentally, the time dilation effect 
due to cosmological redshifts may affect the estimations of timescale, 
but does not 
influence on the logarithmic slopes of structure functions.

\subsection{The starburst model}

The starburst model (SB model) has been proposed 
in order to explain the properties of
radio-quiet AGN, such as broad emission lines and optical variability
(Terlevich et al. 1992; Aretxaga \& Terlevich 1993; Cid Fernandes et al. 1997).
According to this model, the optical variability of AGN can be
understood as the superposition of SNe distributed 
in a Poissonian way (Aretxaga \& Terlevich 1994; Aretxaga 1997; 
Aretxaga, Cid Fernandes \& Terlevich 1997).
In this study, we calculate a fluctuation light curve, 
following Aretxaga et al. (1997).
The generated variability is characterized by the following four parameters:
(1) the background luminosity in {\it B}\,-band due to stars;
       $L_{\it B \, \rm star} \ [L_{\it B \odot}]$,
(2) the time after which radiative cooling becomes important in
the evolution of the compact supernova remnants (cSNRs,
see also Terlevich et al. 1992); $t_{\rm sg}$ [day],
(3) the energy released in {\it B}\,-band by one event; 
$\epsilon_{\it B} \; [10^{51}$ erg], and
(4) the supernova rate;  $\nu_{\rm SN} \ [{\rm yr}^{-1}]$.

The basic element of the light variation is generated
by a SN and the evolution of its associated cSNR.
The shape of luminosity evolution in {\it B}\,-band 
by a single event, $l(t)$, is thus determined by the 
superposition of that of a supernova, 
$l_{\rm SN}$, and that of cSNR, $l_{\rm cSNR}(t)$, where
\begin{eqnarray}
 l_{\rm SN}(t) \ = \ 
 \left\{
  \begin{array}{ll}
	  0   &   {\rm \ for} \qquad t < 0 \\
       6\times 10^{9} \left( 1-{\displaystyle \frac{t}{110} }
       \right) \ [L_{\it B \odot}]  & 
	  {\rm \ for} \qquad 0 \leq t \leq 110 
  \end{array}
 \right.
\end{eqnarray}
\begin{eqnarray}
 l_{\rm cSNR}(t) \ = \ 
 \left\{
  \begin{array}{ll}
   3\times10^{10}\,{\displaystyle\frac{\,365 \ \epsilon_{\it B}\,}{t_{\rm sg}}
   \,\left( \frac{\,t - 0.3 \, t_{\rm sg}\, }{ 0.7 \, t_{\rm sg} } \right)}
   \ [L_{\it B \odot}] & {\rm \ for} \qquad 
      0.3 \, t_{\rm sg} \leq t \leq t_{\rm sg} \medskip \\
   3\times10^{10}\,{\displaystyle\frac{\,365 \ \epsilon_{\it B}\,}{t_{\rm sg}}
   \,\left(\frac{t}{t_{\rm sg}}\right)^{-\frac{11}{7}}} \ [L_{\it B \odot}] &
          {\rm \ for} \qquad t_{\rm sg} \leq t \\
  \end{array}
 \right.
\end{eqnarray}
Therefore, the first peak (at $t=0$) is due to a SN, while
the second one (at $t=t_{\rm sg}$) is due to its cSNR, respectively.
An example of $l(t)$ plotted in Figure 2, 
shows a sudden rise and a gradual decay.  
Here, we use $t_{\rm sg} = 280$ day and 
$\epsilon_{\it B} = 0.5$ ($\times 10^{51}$\,erg).

There are constraints on the four parameters from the observations.
First, since both SN rate ($\nu_{\rm SN}$) and {\it B}\,-band luminosity
coming from stars ($L_{\it B \, \rm star}$) are linked to the number of 
massive stars, the ratio of $\nu_{\rm SN}$ to $L_{\it B \, \rm star}$
is approximately given by the following expression 
(Aretxaga \& Terlevich 1994),
\begin{eqnarray}
\frac{\nu_{\rm SN}}{L_{\it B \, \rm star}} \, \approx \, 
 2 \, \times \, 10^{-11} \,
 [\,{\rm yr}^{-1} \, L_{\it B \odot}^{\, -1}\,].
\end{eqnarray}
Second, the cooling timescale of the cSNRs ($t_{\rm sg}$) was found to 
be $260 \sim 280$ days from the observations of NGC 4151 and NGC 5548 
(Aretxaga \& Terlevich 1993, 1994), 
while for QSOs, which may have higher metallicities 
(Hamann \& Ferland 1992, 1993), cooling timescale could be shorter. 
Therefore, we adopt $t_{\rm sg} \, \lsim \, 280 \, $[day] 
in numerical calculations.
Third, an estimation of $ \epsilon_{\it B} $ can be obtained from the observed 
time averaged equivalent width of H${\beta}$
(for details, see Aretxaga \& Terlevich 1994),
\begin{eqnarray}
\overline{ W_{\rm H{\beta}} } \, \sim \, 320  \,
     \frac{\epsilon_{\it B}}{1 + 0.17 \, \epsilon_{\it B}} \, [{\rm \AA}] .
\end{eqnarray}
With the observed equivalent width of H${\beta}$ in QSOs,
$\overline{ W_{\rm H{\beta}} } \, \sim \, 100 \ {\rm \AA}$
(Osterbrock 1991),
$\epsilon_{\it B} $ is estimated to be $\epsilon_{\it B} \sim 0.5$ 
($\times 10^{51}$ erg).
For the fourth parameter, the SN rate ($\nu_{\rm SN}$) is linked to 
the time average of the total {\it B}\,-band luminosity of the individual AGN 
($\overline{L_{\it B_{\rm T}}}$) according to equation (7) by
\begin{eqnarray}
\overline{L_{\it B_{\rm T}}}\ =\ L_{\it B \, \rm star}\, 
        [L_{\it B \odot}] \ + \ \epsilon_{\it B} \, \nu_{\rm SN}
			\times 10^{51} [{\rm erg} \, {\rm yr}^{-1}] \ 
	  \approx \, 5 \, \times \, 10^{10} \, \nu_{\rm SN} \, 
                  (1 + \epsilon_{\it B} ) \ [L_{\it B \odot}].
\end{eqnarray}

To construct simulated AGN light curves based on this model, 
we superposed SN+cSNR light curves of a given $t_{\rm sg}$ 
at random in time with a rate, $\nu_{\rm SN}$. 
The value of $\nu_{\rm SN}$ determines the event rate and the 
luminosity coming from stars through equation (7), 
while the value of $t_{\rm sg}$ 
determines the cooling timescale and the peak luminosity of cSNR.
Each event may have slightly different cooling timescales 
($t_{\rm sg}$) and released energies ($\epsilon_{\it B}$) around mean values. 
To consider this issue, we assumed the values of 
$t_{\rm sg}$ and $\epsilon_{\it B}$ vary in a Gaussian way around their mean,
such that factor of 2 variation corresponds to
twice the value of the standard  deviation of 
a Gaussian distribution (see Aretxaga et al. 1997).

We performed Monte-Carlo simulations based on the SB model 
described above to calculate light curves and structure functions. 
We focus our attention on the slope of the structure function, $\beta$, 
since this is a key diagnostic to distinguish between the SB model and 
the disk-instability (DI) model.
Figure 3 shows an example of the results: 
the upper two panels represent light curves with different parameter sets,
and the lower panel shows the $[V(\tau)]^{1/2}$ of these light curves
with total time of $8 \times 10^4$ day.
The structure function produced by this model increases
in a power-law (\,$[V(\tau)]^{\frac{1}{2}} \, \propto \, \tau^{\beta}$\,) 
with logarithmic slope ($\beta$) of 0.74 $\sim$ 0.90 
at $\tau \lsim t_{\rm sg}$ depending 
on the parameters (see Table 1), and flattens at the time scale, $t_{\rm sg}$.
The results of all the calculated SB models are summarized in Table 1. 
There is a clear tendency that the slope of the structure function 
depends only on the shape of each shot 
(i.e. $t_{\rm sg}$) and not on the event rate (i.e. $\nu_{\rm SN}$).
This agrees with the theoretical expectation (see \S 2.1).

Incidentally, 0957+561, which we adopted for comparison with models, 
corresponds to $\nu_{\rm SN} \sim$ 50 yr$^{-1}$, 
assuming M$_B \sim$ --26.
The observed light curve yields a more gradual 
slope of the structure function; $\beta \sim 0.35$ (Fig. 1), 
which is too small to account for solely in terms of starbursts.
However, there exists a possibility that thermal instabilities
in cSNRs (e.g. Plewa 1995), which are neglected in the present analytical
approximation [\,Eq. (5), (6)\,], may produce a highly variable light curve
over weeks to months above the light curve calculated by the approximation
(see Cid Fernandes et al. 1996).
If it is the case, 
the resultant slopes for the SB model may change significantly
(Terlevich \& Aretxaga 1998, private communication).

\subsection{The cellular-automaton model for the disk instability}

For the disk-instability model, we consider 
an accretion disk atmosphere emitting 
power-law X-ray spectrum which is substantially fluctuating in time. 
Our hypothesis for generating the optical fluctuation is as follows; 
some instability taking place in the atmosphere 
leads the optical variability. 
According to this hypothesis, we adopt
the cellular-automaton model 
(Mineshige, Ouchi \& Nishimori 1994).
The resultant light curves and PDS produced by the model 
are in good agreement with those of the observed fluctuations.
We pursue their calculation methods to produce fluctuating 
light curves in the optical band, assuming that the optical variability 
simply follows the X-ray variability with little delay.

The procedure to calculate fluctuations is as follows 
(see Takeuchi, Mineshige \& Negoro 1995).

(1) We divide the disk plane into numerous cells along
the two-dimensional circular coordinates ($r, \ \varphi$).
Each cell is thus characterized by two coordinates, $r_i \ 
(i \, = \, 1 - I)$, and $\varphi_j \ (j \, = \, 1 - J)$.
A larger $i$ means a smaller radius ($ r_{i+1} < r_i$ ).

(2) After choosing one cell randomly at the outermost ring (i.e. $i$ = 1),
we put a gas particle with a mass, $m$, into the selected cell.
This process represents a mass supply to
the disk which rotates around a supermassive black hole.

(3) We choose one cell at each ring randomly, and let a small amount of
mass, $m' \ll m$, 
fall into the adjacent inner cell with same $\varphi$.
This process corresponds to gradual viscous diffusion,
and is needed to reproduce the observed properties.

(4) For unstable cells, where a mass density 
exceeds a critical value which is {\it a priori} given, 
we let three mass particles fall from that cell equally
into three adjacent cells at the adjacent inner ring.
In other words, if mass density at ($r_i, \ \varphi_j$) exceeds 
a critical value [i.e. $M_{i,j} > M_{\rm crit}(r_i)$\,], we set 
\begin{eqnarray}
M_{i,j} & \rightarrow & M_{i,j} - 3m ,\nonumber \\
M_{i+1,j \pm 1} & \rightarrow & M_{i+1,j \pm 1} + m ,\\
M_{i+1,j} & \rightarrow & M_{i+1,j} + m .\nonumber
\end{eqnarray}
This process corresponds to an avalanche flow or a flare.
The inner cells may become unstable as the result
of an avalanche flow from above.  
In that case, a subsequent avalanche flow can occur in a next time step 
[after repeating the procedures (2) and (3) above].

(5) We repeat the processes (2) to (4) over $10^4$ times to remove 
the effect of the initial condition and minimize statistical errors.
Each mass blob can travel over one mesh point at
maximum within one time step.
Figure 4 shows a schematic view of those procedure.

It is known (Bak et al. 1988) that under such circumstances the disk 
automaton will evolve to and stay at a self-organized critical state.
In this state, most of single flares calm down without triggering
subsequent avalanches.  
But some single flares trigger
small-scale avalanches over several radial mesh points.
Furthermore, although in quite rare cases, it is possible 
for a single flare to trigger a large-scale avalanche
involving almost the entire region. 
The mass density in each cell ($M_{i,j}$) always
remains slightly lower than the critical 
mass density ($M_{\rm crit}$), thus resultant light curve
does not depend on the value of $M_{\rm crit}$.
Here, we adopt the two dimensional disk 
which undergoes a rigid rotation for simplicity,
but it is easily demonstrated that even if three dimensional structure
of the disk (i.e., $r$-, $\varphi$-, and $z$-directions) and
the effects of differential rotation are taken into account,
the outcome will not change significantly
(Takeuchi, Mineshige \& Negoro 1995).

In the present DI model, the time step is not {\it a priori}\, specified.
To give timescales, we must specify disk models which
describe the dynamical behavior of atmosphere.
For example, if we relate this model to the advection-dominated 
accretion flow (ADAF) model (Abramowicz et al. 1995; Narayan \& Yi 1995; 
Manmoto, Mineshige \& Kusunose 1997), 
the characteristic timescale in the light curve
corresponds to the accretion timescale, $\tau_{\rm acc}$, from 
$r_{\rm out}$ to $r_{\rm in}$ (Manmoto et al. 1996).
The radial velocity of the ADAF model is less than 
or comparable to the free-fall velocity, 
though that of the standard accretion disk model is much less. 
Thus we approximately estimate the timescale ($\tau_{\rm acc}$) 
for the ADAF as
\begin{equation}
 \tau_{\rm acc} \gsim \left(\frac{r^3}{GM} \right)^{1/2}
    = \ 160 \left(\frac{r}{10^2r_{\rm g}}\right)^{3/2}
          \left(\frac{M}{10^9M_\odot}\right)~{\rm day},
\end{equation}
where $r_{\rm g}$ is Schwarzschild radius defined by
\begin{equation}
 r_{\rm g} \equiv \frac{2GM}{c^2}.
\end{equation}

We performed Monte-Carlo simulations for several parameter sets,
following the cellular-automaton procedure.
Examples of the calculated light curves and structure functions 
with a total time step of $8 \times 10^4$ 
are shown in Figure 5.
The model parameters are shown in Table 2; here, $m' = 0.1 \, m$, 
means that the ratio of diffusion mass to inflow mass is $0.1$,
and $r_{\rm in}$ and $r_{\rm out}$ represent the inner radius
and outer radius of the calculated disk region, respectively.
In this model, we found that each $[V(\tau)]^{1/2}$ increases 
in a power-law fashion, 
$[V(\tau)]^{\frac{1}{2}} \, \propto \, \tau^{\beta}$, 
at $\tau \lsim \tau_{\rm acc}$ 
and flattens at the timescale, $\tau_{\rm acc}$, as we have 
seen in the starburst model,
but the logarithmic slopes are systematically smaller 
than those of the SB model;
$\beta = 0.41 \sim 0.49$ (see Table 2), 
with rather weak dependence on the parameters.
These values are closer to the observed ones.
In the present DI model, the resultant logarithmic slopes ($\beta$) 
seem to depend both on the characteristic luminosity profile of 
a largest-scale avalanche and on the distribution of avalanches with 
different scales (\S 2.1).

We do not include the relativistic effects in the present DI model. 
However, the effects are substantial only for gas blobs near the central 
massive black hole, where the dynamical timescale 
$\tau_{\rm dyn}$ is 
$$\tau_{\rm dyn} \sim \left(\frac{r^3}{GM} \right)^{1/2}
    \sim \ 0.8 \left(\frac{r}{3 \, r_{\rm g}}\right)^{3/2}
          \left(\frac{M}{10^9M_\odot}\right)~{\rm day}.$$
Hence they may affect structure functions for 
$\tau \lsim$ 1 day, but cannot substantially 
change the overall shape for $\tau > $ 1 day.
Consequently, we can conclude that 
the DI model is favored over the SB model 
in order to explain a well observed AGN light curve, 
which has the structure function slope of $\beta \sim$ 0.35.

\section{ANALYSIS OF TIME-ASYMMETRY}

Two-point statistics (correlation function, structure function 
or power spectrum) cannot
define an arrow of time from data sets. 
Thus, whether the light curve favors rapid rise (and gradual decay) 
or gradual rise (and rapid decay) can in principle be measured by 
three-point statistical quantities.
To examine such a deviation from time-symmetry,
Press \& Rybicki (1997) simulated light curves by 
multiple superposition with some rate, $\nu$, of a certain 
canonical shot which has a simple time-asymmetric profile. 
They analyzed and compared the light curves of 
0957+561 and the simulated ones in terms of three-point statistics, 
concluding that simulated light curves with 
small $\nu$ have large time-asymmetry, 
and consequently that the observed light curve rules out 
the models with rates
of superposition ($\nu$) that are less than 90 yr$^{-1}$. 

To evaluate the time-asymmetry of the light curve, 
we adopt an alternative approach; we separate $ V(\tau) $ into two parts, 
$V_+(\tau)$ and $V_-(\tau)$ (hereafter plus and minus structure functions,
respectively), depending on the sign of 
[\,$m_{\rm opt}(t_i) - m_{\rm opt}(t_j)$\,];  
\begin{eqnarray}
V_+(\tau) \equiv \frac{1}{N_+(\tau)} 
  \sum_{i<j}\left[\,m_{\rm opt}(t_i)-m_{\rm opt}(t_j)\,\right]^2
  \qquad  \mbox{for} \quad m_{\rm opt}(t_i) - m_{\rm opt}(t_j) > 0, \bigskip \\
V_-(\tau) \equiv \frac{1}{N_-(\tau)} 
  \sum_{i<j}\left[\,m_{\rm opt}(t_i)-m_{\rm opt}(t_j)\,\right]^2
  \qquad  \mbox{for} \quad m_{\rm opt}(t_i) - m_{\rm opt}(t_j) < 0,
\end{eqnarray}
where the summations in the expressions of $V_+(\tau)$ and $V_-(\tau)$
are made, respectively, only for pairs 
which have plus and minus signs of 
$[\,m_{\rm opt}(t_i)-m_{\rm opt}(t_j)\,]$, and
$N_+(\tau)$ and $N_-(\tau)$ are the numbers of such pairs.
Decreasing $m(t_i)$ with time, i.e. 
$m_{\rm opt}(t_i) - m_{\rm opt}(t_j) > 0$, 
represents increasing luminosity with time,
thus $V_+(\tau)$ approximately indicates the structure function 
of brightening phases, and 
similarly $V_-(\tau)$ roughly expresses that of decaying phases.
If the data is produced by time-symmetric processes, 
$V_+(\tau) $ and $V_-(\tau) $ are expected to coincide with $ V(\tau) $.
On the other hand, a significant discrepancy would
indicate a deviation from time-symmetry; 
for example, $V_+(\tau) \gsim V_-(\tau) $ means that 
the light curve favors a rapid rise and gradual decay.

Figure 6 displays the plus and minus structure functions, 
$V_+(\tau)$ and $V_-(\tau)$ (top panel) and 
the relative difference between them, 
$[V_+(\tau)]^{1/2} - [V_-(\tau)]^{1/2}$, 
normalized by the usual structure function 
$[V(\tau)]^{1/2}$ (bottom panel) for the SB model with 
several parameter sets.
Figure (6a) shows the case of $t_{\rm sg} = 85$ day and Figure (6b) is 
for $t_{\rm sg} = 280$ day. 
In both top panels, the plus and minus structure functions 
are represented by upward filled and downward open triangles, respectively.
While the solid lines in all panels represent 
the cases of $\nu_{\rm SN} = 5, \ 20,\ {\rm and}\ 100$ yr$^{-1}$, 
from the top to the bottom respectively.
The error bars in the bottom panels are calculated providing that the 
uncertainty of $[V_+(\tau)]^{1/2} - [V_-(\tau)]^{1/2}$ 
approximately equals to $\sqrt{2}$ times 
the errors in the photometry ($\sim \pm 0.011$ mag).
All the results show time-asymmetry to some extent; 
there is a signature of rapid rise and slow decline,
as expected from the single event profile (figure 2).
As pointed out by Press \& Rybicki (1997), 
there is a trend such that the smaller $\nu_{\rm SN}$ is, 
the larger the resulting time-asymmetry in the SB model.
Within errors in the photometry, both of 
$V_+(\tau)$ and $V_-(\tau)$ are consistent with time-symmetry 
except for the models with
$\nu_{\rm SN} \ \lsim$ 10 yr$^{-1}$, which shows a significant 
discrepancy at $\tau \ \lsim$ 100 day.

We also calculated the plus and minus structure functions for the 
DI model, which are depicted in figure 7.
Here the solid lines in the bottom panel represent
the cases of 
$m' = 0.5 \, m, \ 0.1 \, m \ {\rm and} \ 0.02 \, m$, 
from the top to the bottom respectively.
In contrast to the SB model,
the relative difference between them shows the signature
of a gradual rise and a rapid decay, and moreover, 
the tendency becomes enhanced in the case of $m' \lsim 0.1 \, m$.
This time-asymmetry comes from
the shape of the luminosity variation due to a single 
large-scale avalanche.
An avalanche flow starting at some outer radius leads to luminosity variation
over an accretion timescale, $\tau_{\rm acc}$,
in which the emissivity increases as the flow approaches the center.
Thus the light curve shows a gradual brightening over $\tau_{\rm acc}$.
After the front of the flow reaches the innermost region,
the amount of radiation rapidly decays, 
which is responsible for the rapid decline in the light curve
(see, however, section 4 for hydrodynamic effects).
However, a large $m'$ causes frequent small-scale avalanches, 
which hide the time-asymmetry.

Finally, Figure 8 displays $V_+(\tau)$ and $V_-(\tau)$ and their 
relative differences 
for the observed light curve of 0957+561.
The bottom panel of the figure shows the relative differences fluctuating at 
$\tau \ \gsim$ 70 day, which can be ascribed to the effect 
of the finite length of the observational light curve.
Thus we focus our attention solely to the differences
at $\tau \ \lsim$ 70 day.
In this range of time lags $\tau$,
apparent time-asymmetry appears 
in the sense of a slow rise and a rapid decline,
but it is premature to conclude that this tendency is real. 
This is because photometric monitoring over two years is 
still too short to calculate reliable three-point statistics.  
Longer observational data sets will 
be valuable for this purpose
[\,see also Press \& Rybicki (1997)\,].
Further, estimating time-asymmetry from longer observations
will pose effective constraints on models even if they 
possess similar logarithmic slopes ($\beta$) in the structure functions.

\section{DISCUSSION}

We have compared two models, DI and SB, to see how well 
the statistical properties of the best available AGN optical 
light curve can be understood.
In this section, we will discuss several related issues.

The first issue concerns 
the time-asymmetry in simulated light curves based on the two models.
Light curves in the DI model obtained by the present cellular-automaton
simulations show an apparent time-asymmetry with 
slow rise and rapid decline 
under the condition $m' \lsim 0.1 \, m$.
This is because each shot profile obtained by the model possesses 
a time-asymmetry and this becomes more enhanced
in the case of smaller  $m'$, as we mentioned above.
We should be, however, careful about this result, 
since the cellular-automaton simulations neglect hydrodynamic effects.
According to the one-dimensional, hydrodynamic simulations 
of X-ray fluctuations from black-hole objects by Manmoto et al. (1996), 
an individual shot has a rather time symmetric profile
due to wave reflection (see also Takeuchi \& Mineshige 1997).
Such hydrodynamic effects as wave reflection are not included 
in the present simulations.
If they are included, the light curves would show
more time-symmetry.
The SB model, on the other hand, predicts clear time-asymmetry 
(rapid rise and slow decline), unless 
$\nu_{\rm SN} \ \gsim$ 10 yr$^{-1}$.
In the context of the SB model 
such a high supernova rate implies 
objects of larger luminosity such as QSOs,
thus time-asymmetry of such luminous AGN might
be hidden by the observational errors.

Second, we consider the wavelength dependence of 
the timescales of AGN optical variability.
There is a discussion of the origin of AGN long-term variability
initiated by Hawkins (1993), who has claimed that gravitational 
microlensing can account for the absence of the expected time dilation
effect in the observations of high redshift quasar.
Alternatively, Baganoff \& Malkan (1995) have suggested
that the relatively hot region of a standard accretion disk 
radiating at shorter wavelengths should be restricted to the 
inner region.
Thus, when we observe objects with higher redshift, we detect radiation of 
shorter wavelengths in the rest frame, i.e., those coming from the 
inner part, where temperatures are higher and dynamical timescales 
are shorter.
This roughly cancels out the expected time dilation effect.
Hawkins \& Taylor (1997), however, argue that the standard accretion 
disk model predicts wavelength dependent variability timescales as stated
in Baganoff \& Malkan (1995) and that the prediction is inconsistent with 
the observations of the quasar sample and NGC5548.
They conclude that microlensing is preferable
as an explanation for quasar variability, 
although the nearby Seyfert galaxies are probably 
showing intrinsic variations.

In relation to this issue, we can make two points. 
Contrary to the claim by Hawkins \& Taylor (1997), 
microlensing of an optically thick standard accretion disk 
produces wavelength dependent time variations
(Yonehara et al. 1997a).
Furthermore, if occasional avalanche flows occur 
at various radii, 
as we have found in the cellular automata DI model, wavelength 
independent variation timescales are produced naturally 
(Yonehara et al. 1997b).

Third, we discuss the amplitudes of the luminosity variations 
as a function of wavelength.
Multi-wavelength observations of AGNs have revealed that the 
variation amplitude in optical -- ultraviolet bands is 
larger at shorter wavelengths
(Di Clemente et al. 1996; Edelson et al. 1996).
This suggests that the outer cool region of the disk, emitting at 
longer wavelengths, is less variable than the inner hot region, emitting at 
shorter wavelengths.
This tendency could also be understood in our
DI model, since an avalanche flow starting at the outer region drifts 
down spreading the avalanche region (see also Yonehara et al. 1997a),
to a larger fraction of the inner region.

Fourth, we briefly comment on the relation between 
X-ray variability of AGNs and that of X-ray binaries (XBs).
X-rays from AGNs and XBs are known to behave in 
the similar manner and to exhibit similar shapes of PDS and similar 
shot-like features in their light curves.
This similarity strongly suggests a common mechanism being 
responsible for the X-ray variability of both types of objects.
It is obvious that XB variability cannot be explained by 
microlensing or starburst models, but there is certainly 
evidence that it is associated with an accretion disk.
If AGN variability and XB variability do have the same origin,
as strongly suggested by their similar statistical properties 
(Hayashida et al. 1997), AGN variability is also likely to be 
of accretion disk origin.

Finally, there still remains, of course, a possibility that 
AGN variability is caused by several independent mechanisms. 
The composite effects of the different variability mechanisms on 
the structure functions may be investigated in the future.

\section{CONCLUSION}

We have calculated the structure functions $V(\tau)$ for 
simulated light curves of the two models and the observed 0957+561
light curve to determine which provides a better match.

(1) We summarize the values of the logarithmic slopes $\beta$ 
of $V(\tau)$ among the two models and the observation;
$\beta \sim$ 0.74--0.90 in the SB model and 
$\beta \sim$ 0.41--0.49 in the DI model, 
while the observed light curve exhibits $\beta \sim 0.35$.
Therefore we may conclude the DI model is preferable to the SB model 
to account for the observed value of $\beta$, in the case of 0957+561.
There exists, however, a possibility that the SB model might
yield a more gradual slopes, if thermal instabilities 
in supernova remnants were incorporated.

(2) The two models exhibit opposite trends of time-asymmetry,
though the size of the deviations from time-symmetry depend on 
the model parameters.
Thus, the time-asymmetry in the light curves potentially offer
effective information to test the models.
However, the presently available observational data sets do not 
allow us to make use of this diagnostic.
Thus, we must await longer duration observational light curves.

\acknowledgements

We acknowledge Roberto Terlevich and Itziar Aretxaga for valuable 
comments on the starburst model, and the anonymous referee for 
criticisms that helped improve the paper.
One of the authors (T. K.) wishes to thank Karen M. Leighly for 
useful discussions regarding time series analysis.
He also thanks Tsutomu T. Takeuchi, Atsunori Yonehara, 
and Hiroyuki Hirashita for helpful comments and encouragement.
This work was supported in part by the Japan-US Cooperative
Research Program which is founded by the Japan Society for
the Promotion of Science and the US National Science
Foundation, and by the Grants-in Aid of the
Ministry of Education, Science, and Culture of Japan,
08640329, 09223212 (S.M.).

\clearpage

\begin{table}
\begin{center}
\tabcolsep=8mm
\begin{tabular}{cccc}
\tableline
\tableline
  $\nu_{\rm SN} \ [{\rm yr}^{-1}]$  &      & $t_{\rm sg} $ [day] &       \\
\cline{2-4}
    		          &  85  &      180         &  280  \\
\tableline
                   5	  & 0.90 &     0.82         & 0.74  \\
		  20	  & 0.90 &     0.83         & 0.74  \\
                 100 	  & 0.90 &     0.83         & 0.74  \\
\tableline
\end{tabular}
\end{center}
\caption{The slopes of the structure function, $\beta$, for the 
starburst model, where we assumed 
$[V(\tau)]^{\frac{1}{2}} \propto \tau^{\beta}$.
Here, $\nu_{\rm SN}$ is supernova rate and
$t_{\rm sg}$ represents the cooling timescale of cSNR.
Since the second parameter, $t_{\rm sg}$, actually 
controls the shape of one shot, resultant slopes of structure functions 
depend on $t_{\rm sg}$ solely (see \S 2.1).}
\end{table}

\clearpage

\begin{table}
\begin{center}
\tabcolsep=8mm
\begin{tabular}{cccc}
\tableline
\tableline
  $r_{\rm in}:r_{\rm out}$  &       &  $m' \, / \, m$	 &       \\
\cline{2-4}
    		      &  0.02 &  0.1   &  0.5  \\
\tableline
        1 : 4         &  0.49 &  0.46  & 0.44  \\
	1 : 20	      &  0.43 &  0.42  & 0.41  \\
  	1 : 100	      &  0.42 &  0.42  & 0.41  \\
\tableline
\end{tabular}
\end{center}
\caption{Same as Table 1 but for the disk-instability model.
Here, $m' = 0.1 \, m$ means that the ratio of diffusion mass to inflow 
mass is $0.1$, and $r_{\rm in}$ and $r_{\rm out}$ denote
the inner radius
and the outer radius of the calculated disk region, respectively.}
\end{table}

\clearpage

\begin{figure}
\caption{The light curve and structure function of 0957+561 
calculated from the observational data by Kundi\'{c} et al. (1997).}
\end{figure}

\begin{figure}
\caption{A typical light curve of one shot in the starburst model for
$t_{\rm sg}$ = 280 day and $\epsilon_{\it B}$ = 0.5 ($\times 10^{51}$erg).
The first peak (at $t = 0$) is due to a SN
and the second peak (at $t=t_{\rm sg}$) is caused by a subsequent cSNR.
The light variation is time-asymmetric; rapid rise and gradual decay 
are shown.}
\end{figure}

\begin{figure}
\caption{Example of the light curves (the upper two panels)
and the structure functions (the bottom panel) produced by the 
starburst model. 
In this figure, $\nu_{\rm SN}$ = 20 ${\rm yr}^{-1}$ (the rate of SNe),
and $t_{\rm sg}$ (the cooling timescales of cSNRs) are labeled inside 
each panel.
The structure functions ($[V(\tau)]^{\frac{1}{2}}$) grows in a power-law way
($[V(\tau)]^{\frac{1}{2}} \, \propto \, \tau^{\beta}$) at small $\tau$.}
\end{figure}

\begin{figure}
\caption{A schematic view of our cellular-automaton model.}
\end{figure}

\begin{figure}
\caption{Same as Fig. 3 but for our cellular-automaton model 
for the disk instability.
Here, $r_{\rm in}:r_{\rm out}$=1:20 (the ratio 
of inner edge to outer edge of the disk) and $m'/m$
(the ratio of diffusion mass to inflow mass) is labeled in each panel.}
\end{figure}

\clearpage
\begin{figure}
\caption{Time-asymmetry calculated 
in terms of the plus and minus structure functions, 
$[V_+(\tau)]^{1/2}$ and $[V_-(\tau)]^{1/2}$, for the SB model
with $t_{\rm sg}=85$ day (6a) and with $t_{\rm sg}=280$ day (6b),
respectively.
Top panels; the usual structure functions are shown by the solid lines,
while the plus and minus structure functions are indicated by the symbols,
upward filled and downward open triangles, respectively.
See equations (13) and (14) for their definitions.
The models show time-asymmetry to some extent in the sense of
a rapid rise and a slow decline.
Bottom panels; the relative differences between 
$[V_+(\tau)]^{1/2}$ and $[V_-(\tau)]^{1/2}$ normalized by 
$[V(\tau)]^{1/2}$ are depicted.}
\end{figure}

\begin{figure}
\caption{Same as figure 6 but for the DI model
with $m'/m=$ 0.02, 0.1, and 0.5.  The ratio of the outer to inner
disk radii is fixed to be $r_{\rm out}/r_{\rm in}=20$.
The models show some time-asymmetry in the sense of
a slow rise and a rapid decline.}
\end{figure}

\begin{figure}
\caption{The plus and minus structure functions calculated from 
the observations of 0957+561, corresponding to figures 6 and 7. 
There is a hint of a tendency for slow rise and rapid decay.}
\end{figure}



\begin{thebibliography}{}

\bibitem[]{}
 Abramowicz, M. A., Chen, X., Kato, S., Lasota, J. -P., \& Regev, O. 1995,
 ApJ, 438, L37
\bibitem[]{}
 Aretxaga, I. 1997, Review of Mexican Astronomy and Astrophysics, {6}, 207
\bibitem[]{}
 Aretxaga, I., Cid Fernandes, R., \& Terlevich R. J. 1997, MNRAS, {286}, 271
\bibitem[]{}
 Aretxaga, I., Terlevich R. 1993, ApSS, {205}, 69
\bibitem[]{}
 Aretxaga, I., Terlevich R. 1994, MNRAS, {269}, 462
\bibitem[]{}
 Baganoff, F., K., \& Malkan, M., A. 1995, ApJ, {444}, L13
\bibitem[]{}
 Bak, P., Tang C., Wiesenfeld K. 1988, Phys. Rev., A38, 364
\bibitem[]{}
 Cid Fernandes, R., Terlevich, R., Aretxaga, I. 1997, MNRAS, {289}, 318
\bibitem[]{}
 Cid Fernandes, R., Plewa, T., R\'{o}\.{z}yczka, M., Franco, J., 
Terlevich, R., Tenorio-Tagle, G., \& Miller, W. 1996, MNRAS, {283}, 419
\bibitem[]{}
 Clavel, J. C. et al. 1992, ApJ, {393}, 113
\bibitem[]{}
 Cristiani, S., Trentini, S., La Franca, F., Aretxaga, I., Andreani, P., 
 Vio, R., \& Gemmoo, A. 1996, AA, {306}, 395
\bibitem[]{}
 Di Clemente, A., Giallongo, E., Natali, G., Tr\`{e}vese, D., \& Vagnetti, F.,
 1996, ApJ, {463}, 466
\bibitem[]{}
 Edelson, R. A. et al. 1996, ApJ, {470}, 364
\bibitem[]{}
 Green, A. R., McHardy, I. M., \& Lehto, H. J. 1993, MNRAS, {265}, 664
\bibitem[]{}
 Haarsma, D. B., Hewitt, J. N., Lehar, J., Burke, B. F. 1997, {479}, 102
\bibitem[]{}
 Hamann, F., Ferland, G. 1992, ApJ, {391}, L53
\bibitem[]{}
 Hamann, F., Ferland, G. 1993, ApJ, {418}, 11
\bibitem[]{}
 Hawkins, M., R., S. 1993, Nature, {366}, 242
\bibitem[]{}
 Hawkins, M., R., S., \& Taylor, A., N. 1997, ApJ, {482}, L5
\bibitem[]{}
 Hayashida, K., Miyamoto, S., Kitamoto, S., \& Inoue, H. 1997, ApJ, submitted
\bibitem[]{}
 Hughes, P. A., Aller, H. D., \& Aller, M. F. 1992, ApJ, {396}, 469
\bibitem[]{}
 Hook, I. M., McMahon, R. G., Boyle, B. J., \& Irwin, M. J. 1994, MNRAS, {268}, 305
\bibitem[]{}
 Krolik, J. H., Kallman, T. R., Malkan, M. A., Edelson, R. A., \& Kriss, G. A. 1991, ApJ, 371, 541
\bibitem[]{}
 Kundi\'{c}, T. et al. 1995, ApJ, 455, L5
\bibitem[]{}
 Kundi\'{c}, T. et al. 1997, ApJ, 482, 75
\bibitem[]{}
 Lawrence, A. \& Papadakis, I. 1993, ApJ, 414, L85
\bibitem[]{}
 Leighly, K. M., \& O'Brien, P. T. 1997, ApJ, 481, L15
\bibitem[]{}
 Malkan, M., A. 1983, ApJ, {268}, 582
\bibitem[]{}
 Manmoto, T., Mineshige, S., Kusunose, M. 1997, ApJ, {489}, 791
\bibitem[]{}
 Manmoto, T., Takeuchi, M., Mineshige, S., Matsumoto, R., and Negoro, H.
 1996, ApJ, 464, L135
\bibitem[]{}
 Mineshige, S., Ouchi, B. N., \& Nishimori, H. 1994, PASJ, 46, 97
\bibitem[]{}
 Narayan, R., Yi, I. 1995, ApJ, 452, 710
\bibitem[]{}
 Netzer, H., \& Peterson, B. M. 1997, in Astronomical Time Series,
 D. Maoz, A. Stermberg, and E. M. Leibowitz (eds.), Dordrecht Kluwer, p.85
\bibitem[]{}
 O'Brien, P. T., \& Leighly, K. M. astro-ph/9701105
\bibitem[]{}
 Osterbrock, D. E., 1991, Rep. Prog. Phys., {54}, 579
\bibitem[]{}
 Peterson, B. M. et al. 1994, ApJ, {425}, 622
\bibitem[]{}
 Plewa, T. MNRAS, {275}, 143
\bibitem[]{}
 Press, W. H., 1978, Comments Ap., {7}, 103
\bibitem[]{}
 Press, W. H., Rybicki, G. B., \& Hewitt, J. N. 1992, ApJ, 385, 404
\bibitem[]{}
 Press, W. H., \& Rybicki, G. B. 1997, in Astronomical Time Series,
 D. Maoz, A. Stermberg, and E. M. Leibowitz (eds.), Dordrecht Kluwer, p.61
\bibitem[]{}
 Rees, M. J. 1984, ARA\&A, {22}, 471
\bibitem[]{}
 Santos-Lle\'{o}, M., Clavel, J., Barr, P. Glass, I. S., Pelat, P., Peterson, B. M., \& Reichert, G. 1995, MNRAS, {274}, 1
\bibitem[]{}
 Schild, R. E. 1996, ApJ, {464}, 125
\bibitem[]{}
 Shields, G., A. 1978, Nature, {272}, 706
\bibitem[]{}
 Simonetti, J. H., Cordes, J. M., \& Heeschen, D. S. 1985, ApJ, 296, 46
\bibitem[]{}
 Takeuchi, M., \& Mineshige, S. 1997, ApJ, 486, in press
\bibitem[]{}
 Takeuchi, M., Mineshige, S., \& Negoro, H. 1995, PASJ, 47, 617
\bibitem[]{}
 Tanaka, Y. et al. 1995, Nature, {375}, 659
\bibitem[]{}
 Terlevich, R., Tenorio-Tagle, G., Franco, J., \& Melnick, J. 1992, MNRAS, 
 {255}, 713
\bibitem[]{}
 Tr\`{e}vese, D., Kron, R. G., Majewski, S. R., Bershady, M. A., \& Koo, D. C. 
  1994, ApJ, {433}, 494
\bibitem[]{}
 Warwick, R. S. et al. 1996, ApJ, {470}, 349
\bibitem[]{}
 Yonehara, A., Mineshige, S., Fukue, J., Umemura, M., \& Turner, E., L.
  1997a, astro-ph/9710190
\bibitem[]{}
 Yonehara, A., Mineshige, S., \& Welsh, W., F. 1997b, ApJ, {486}, 388
\end{thebibliography}
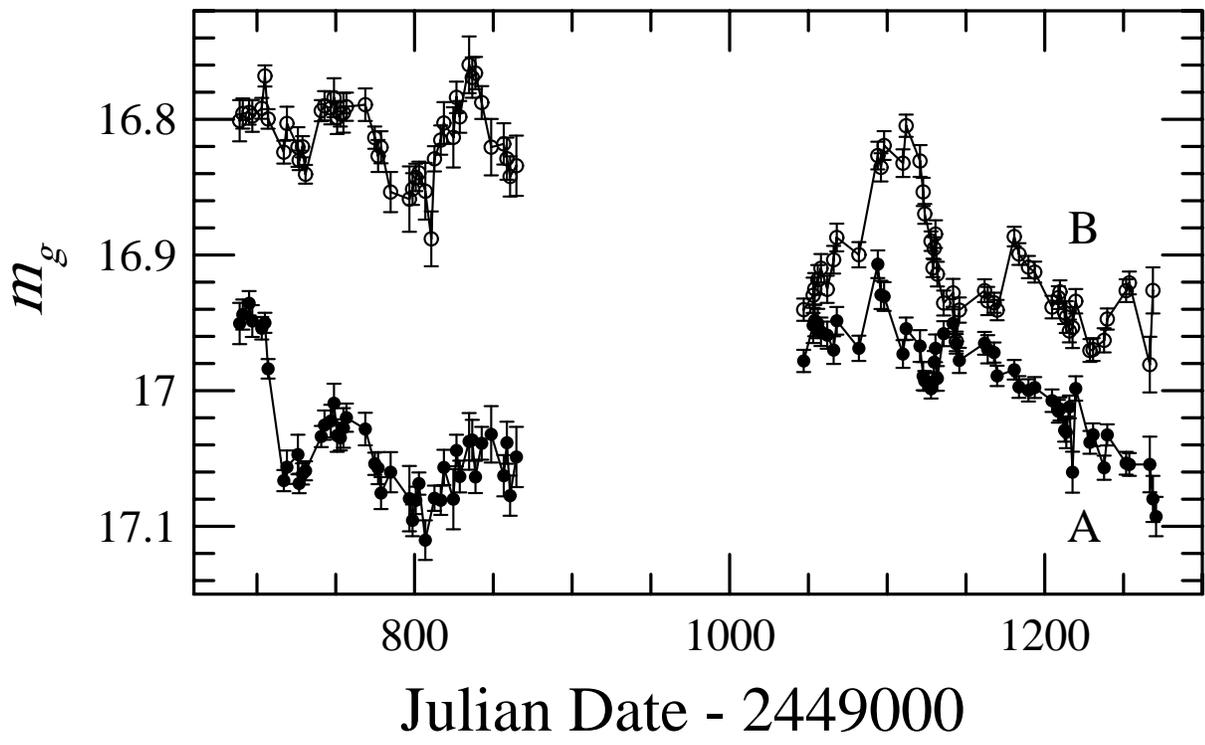
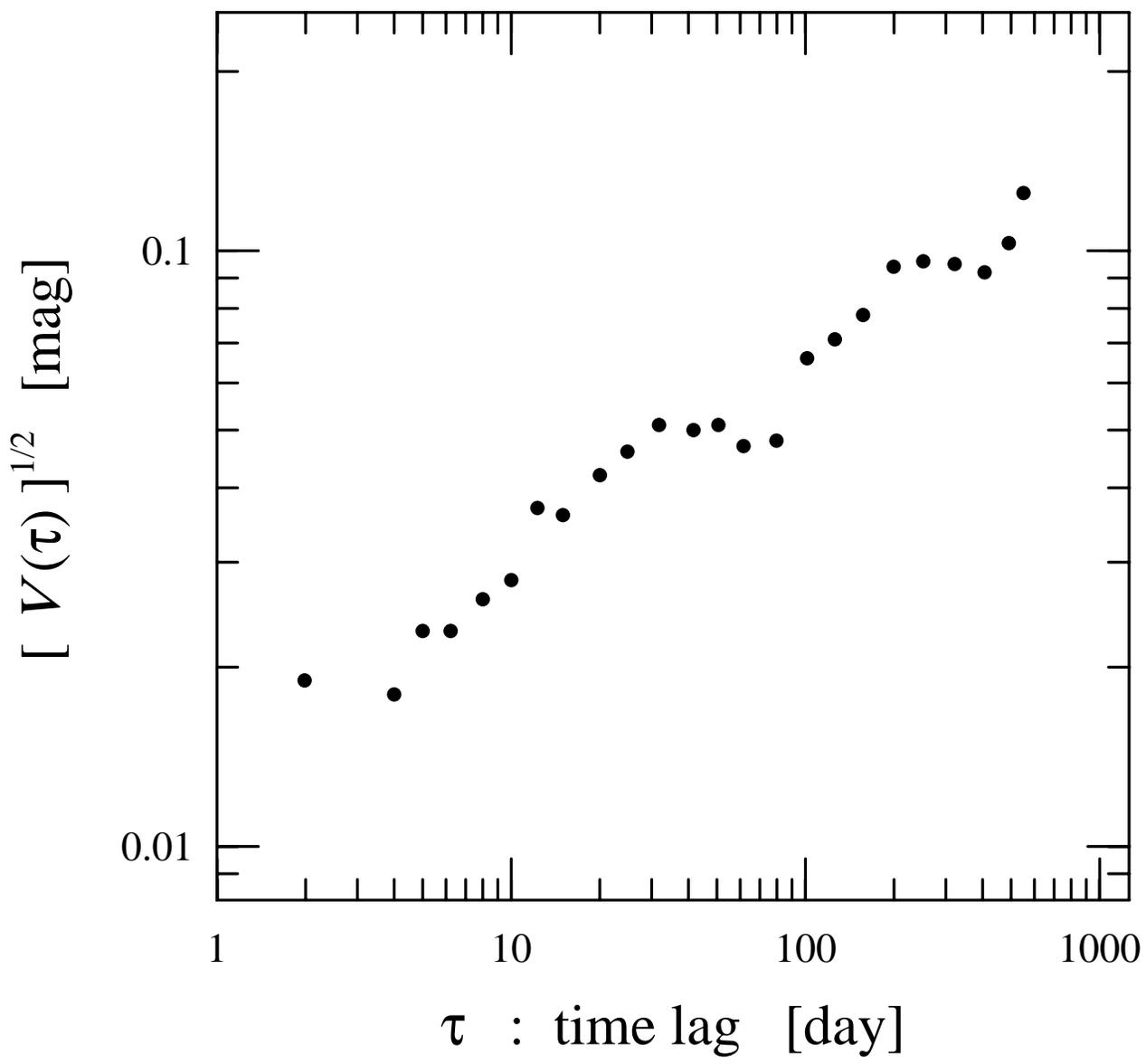
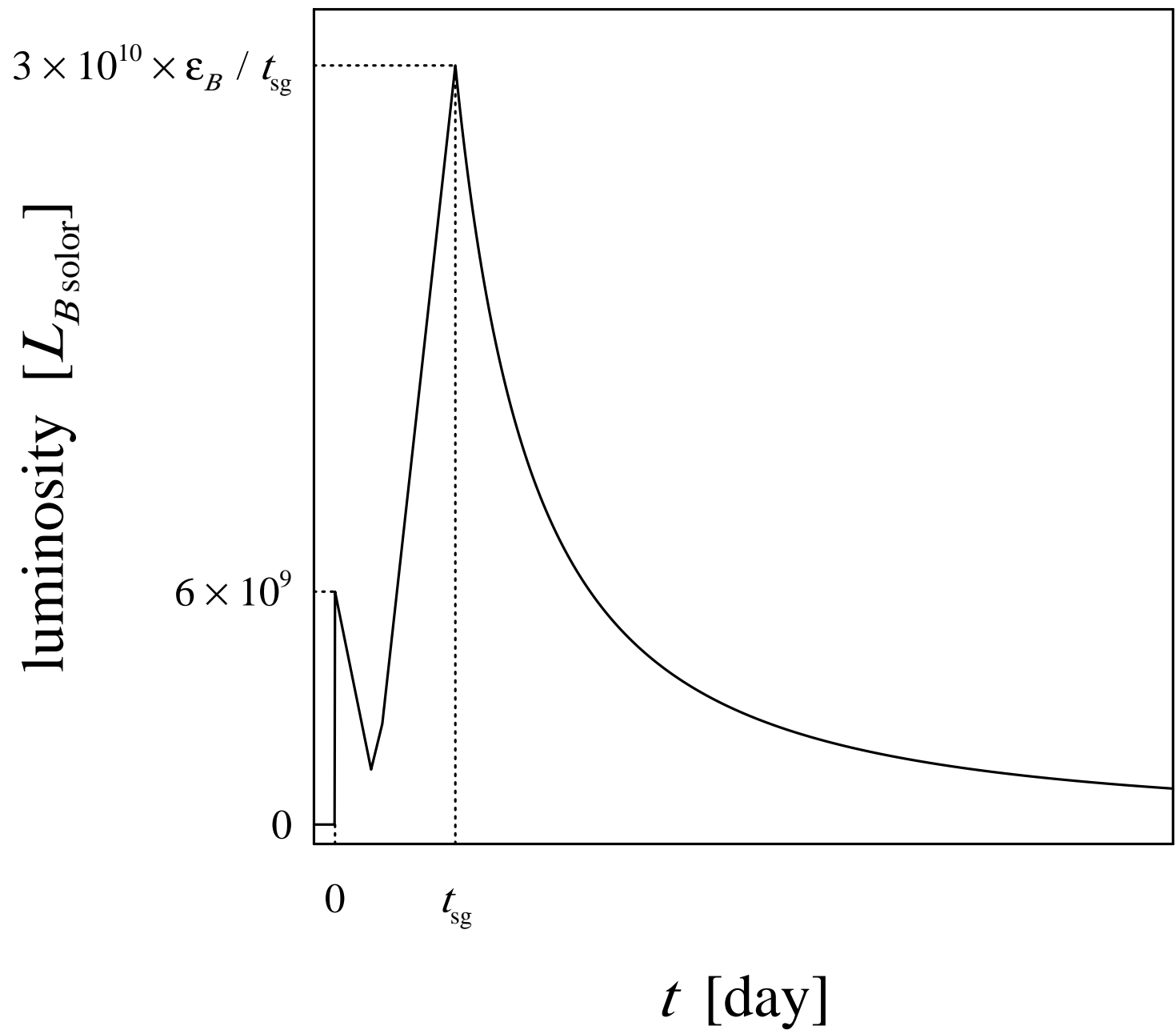
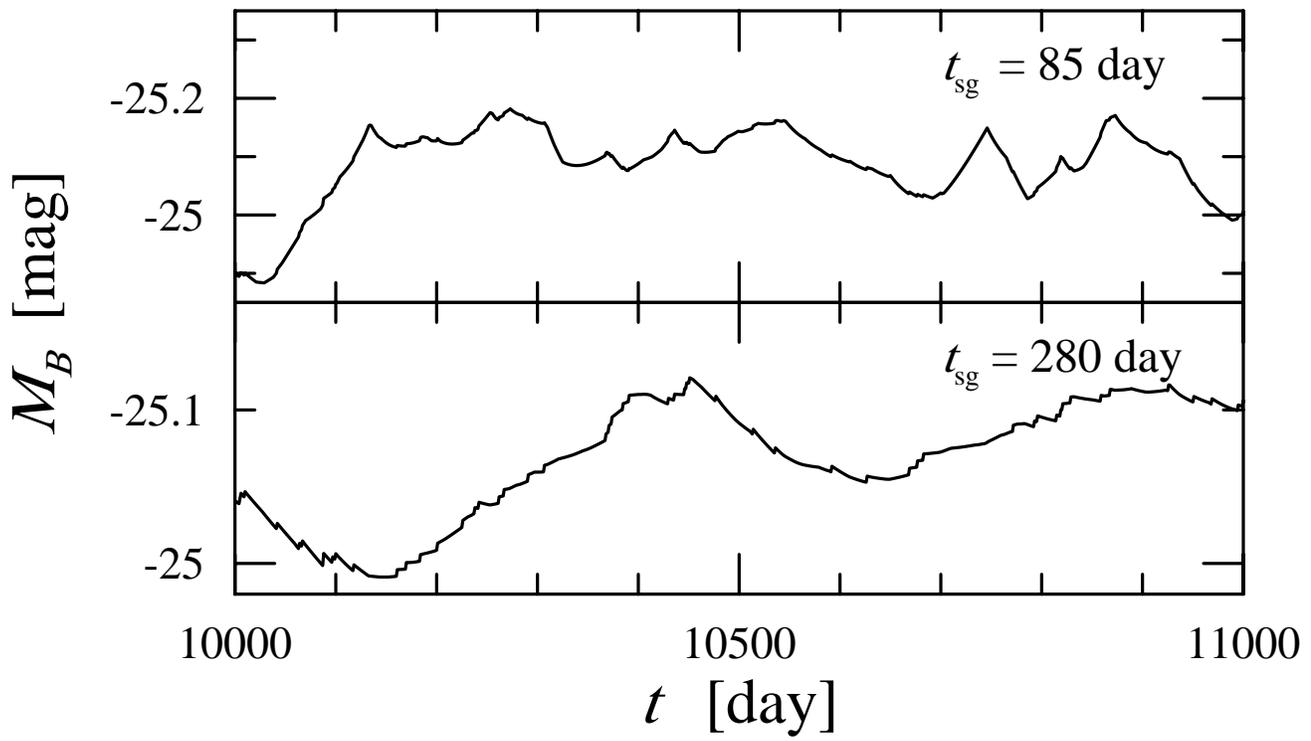
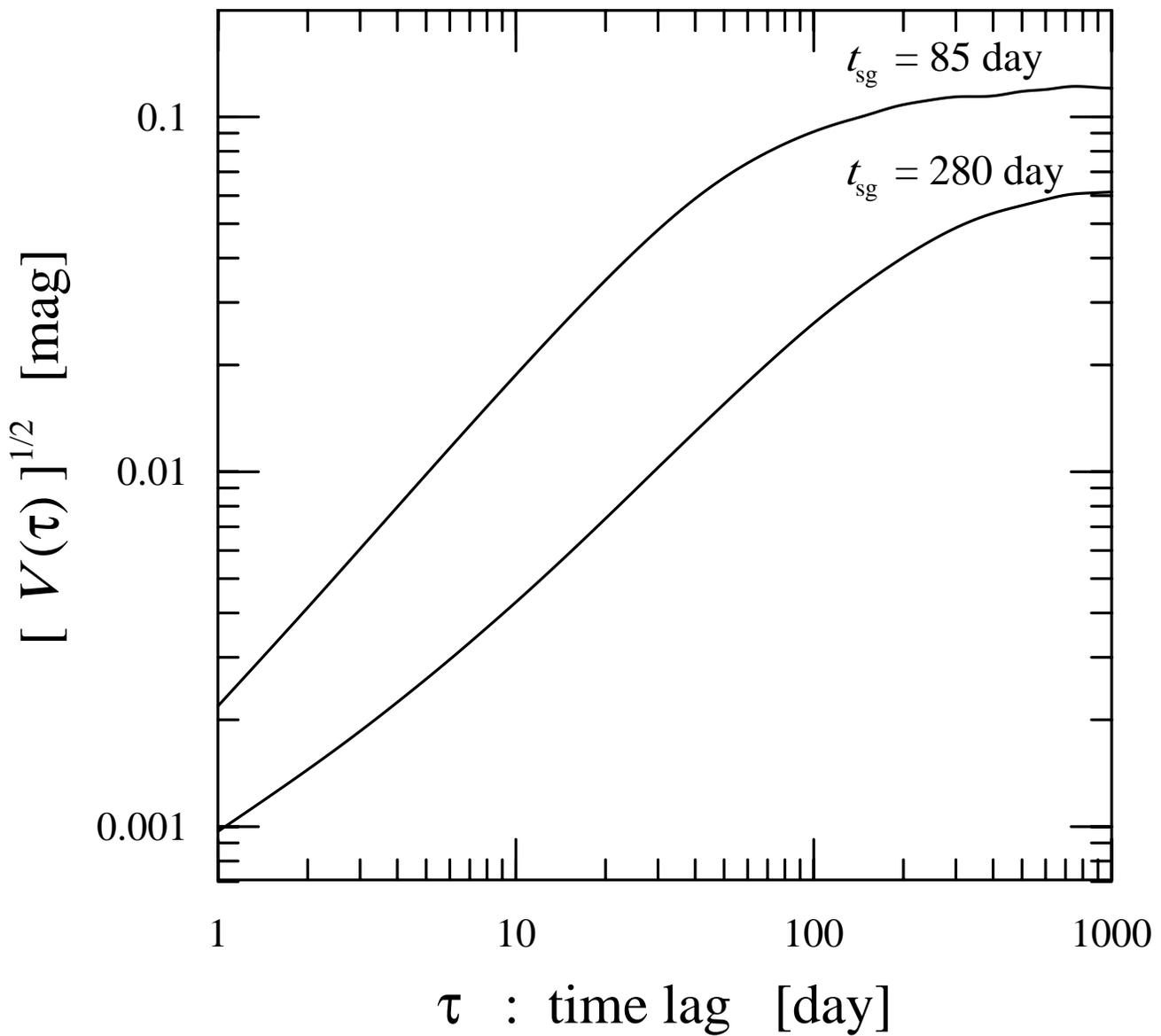
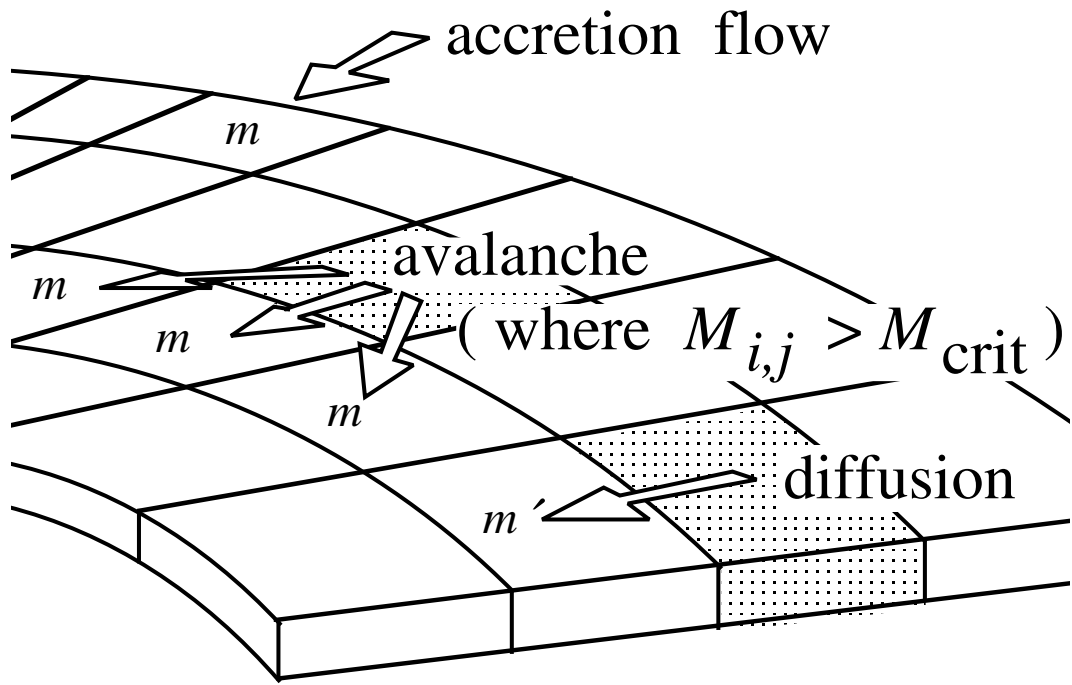
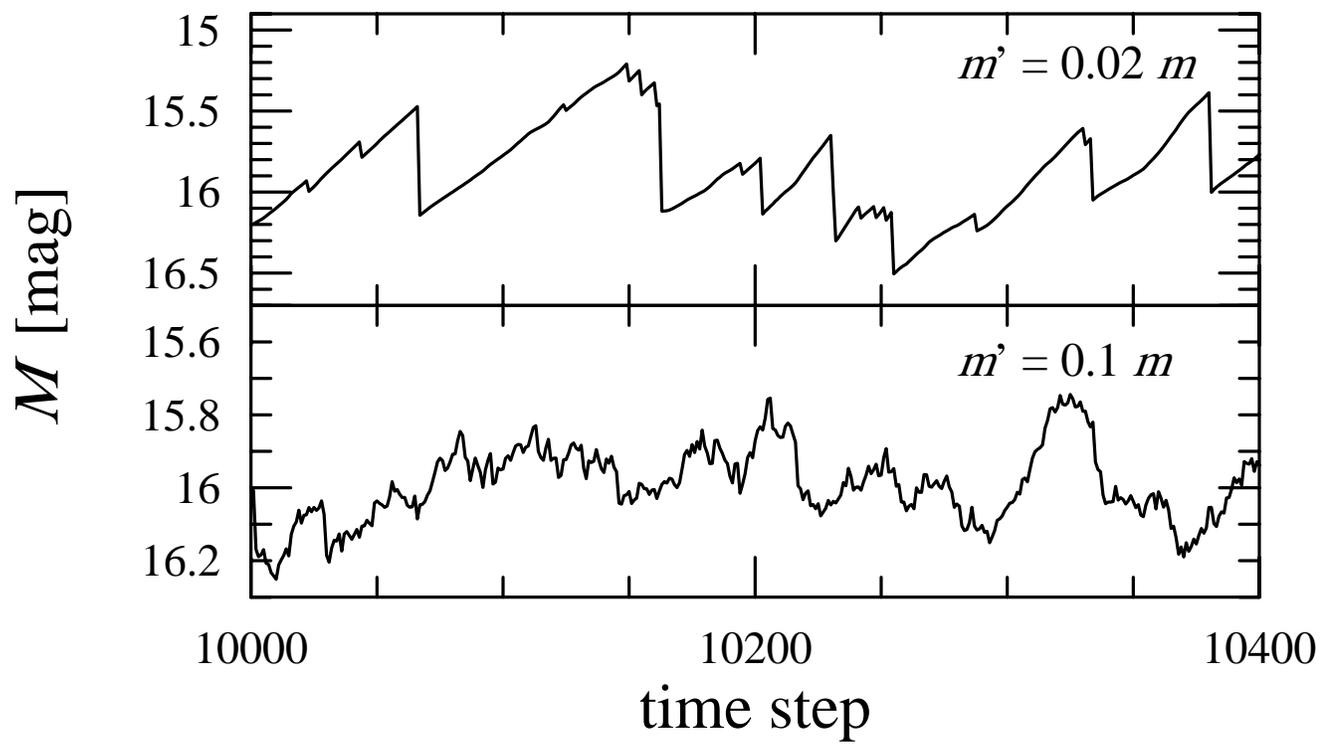
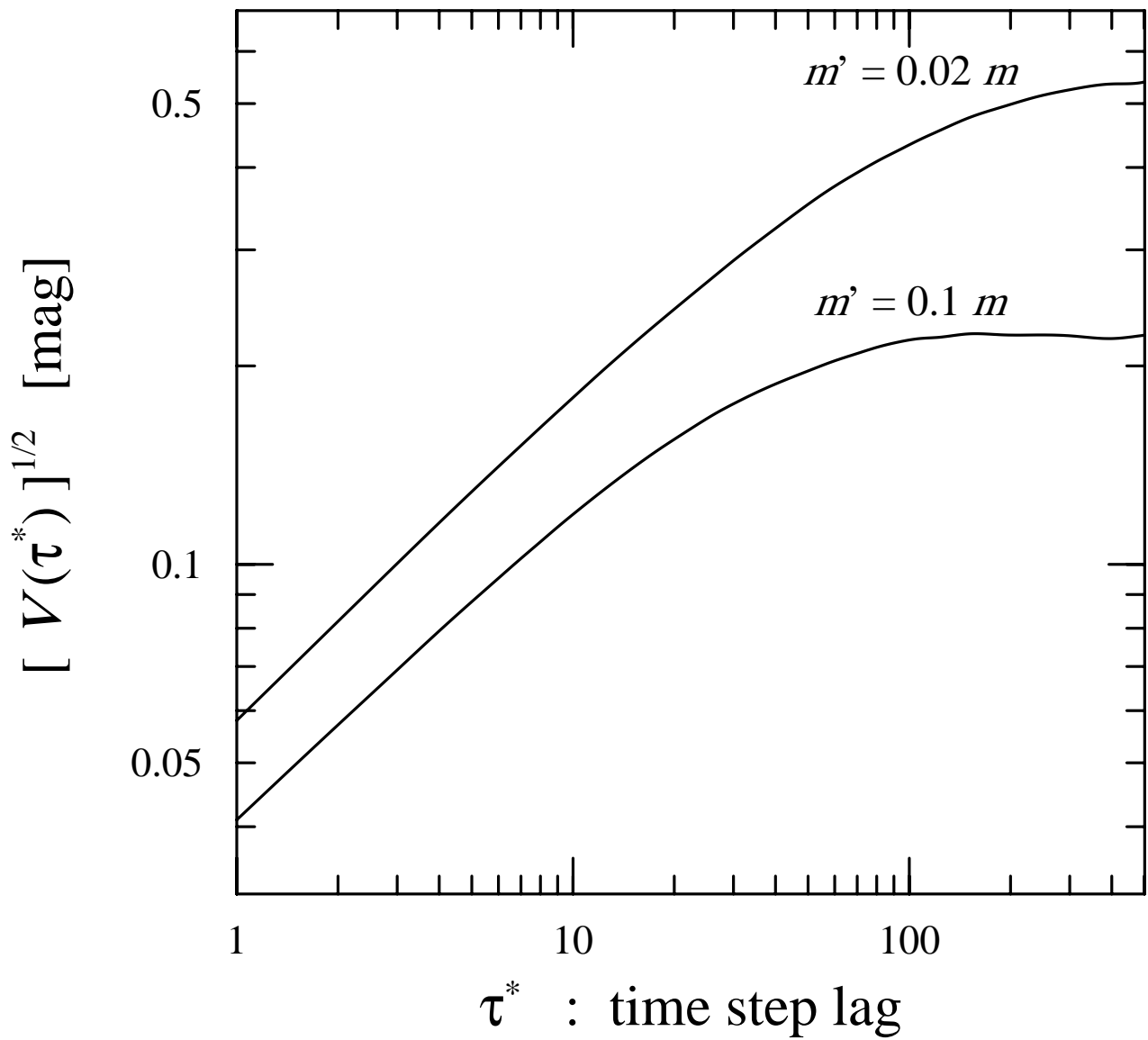
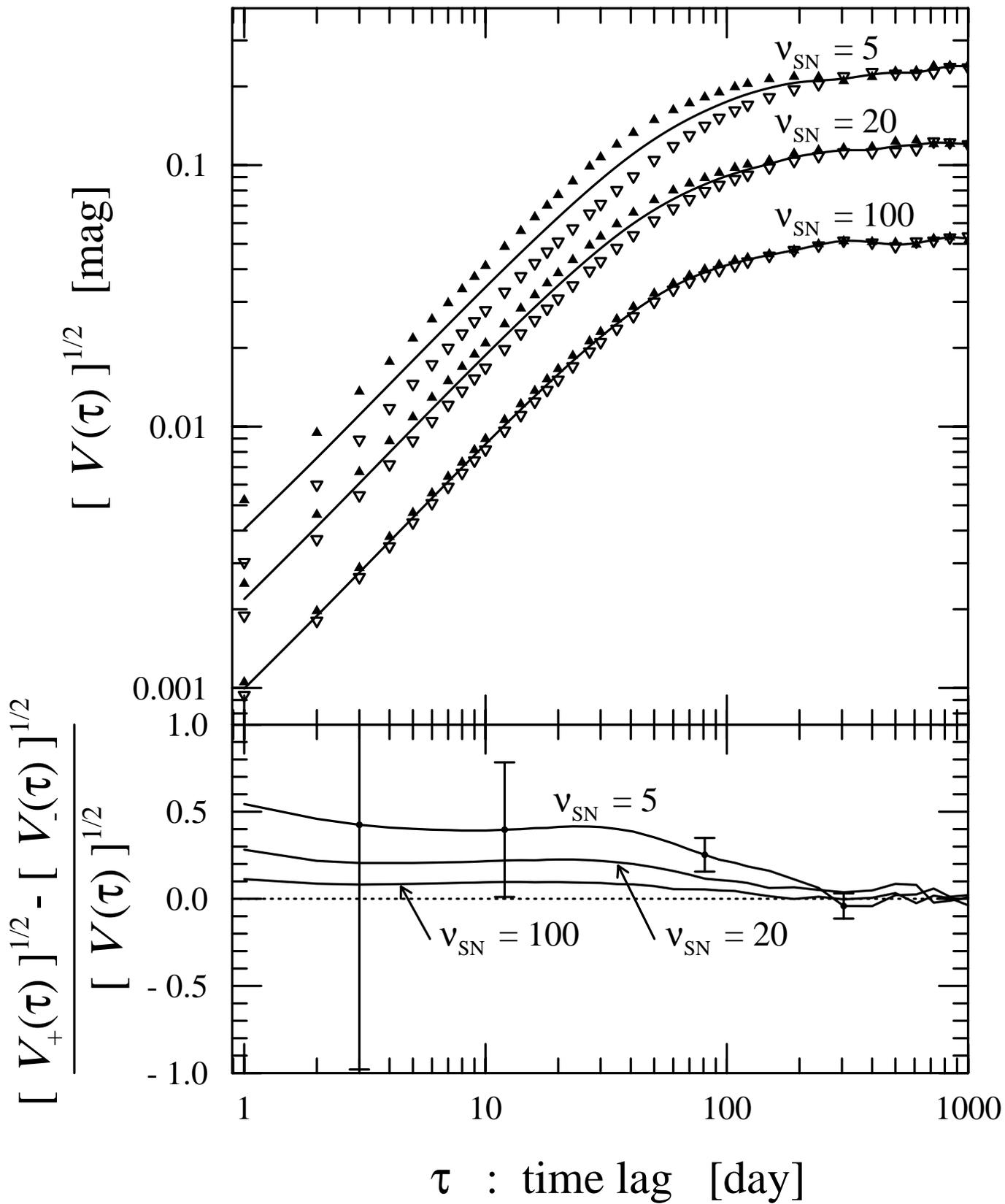
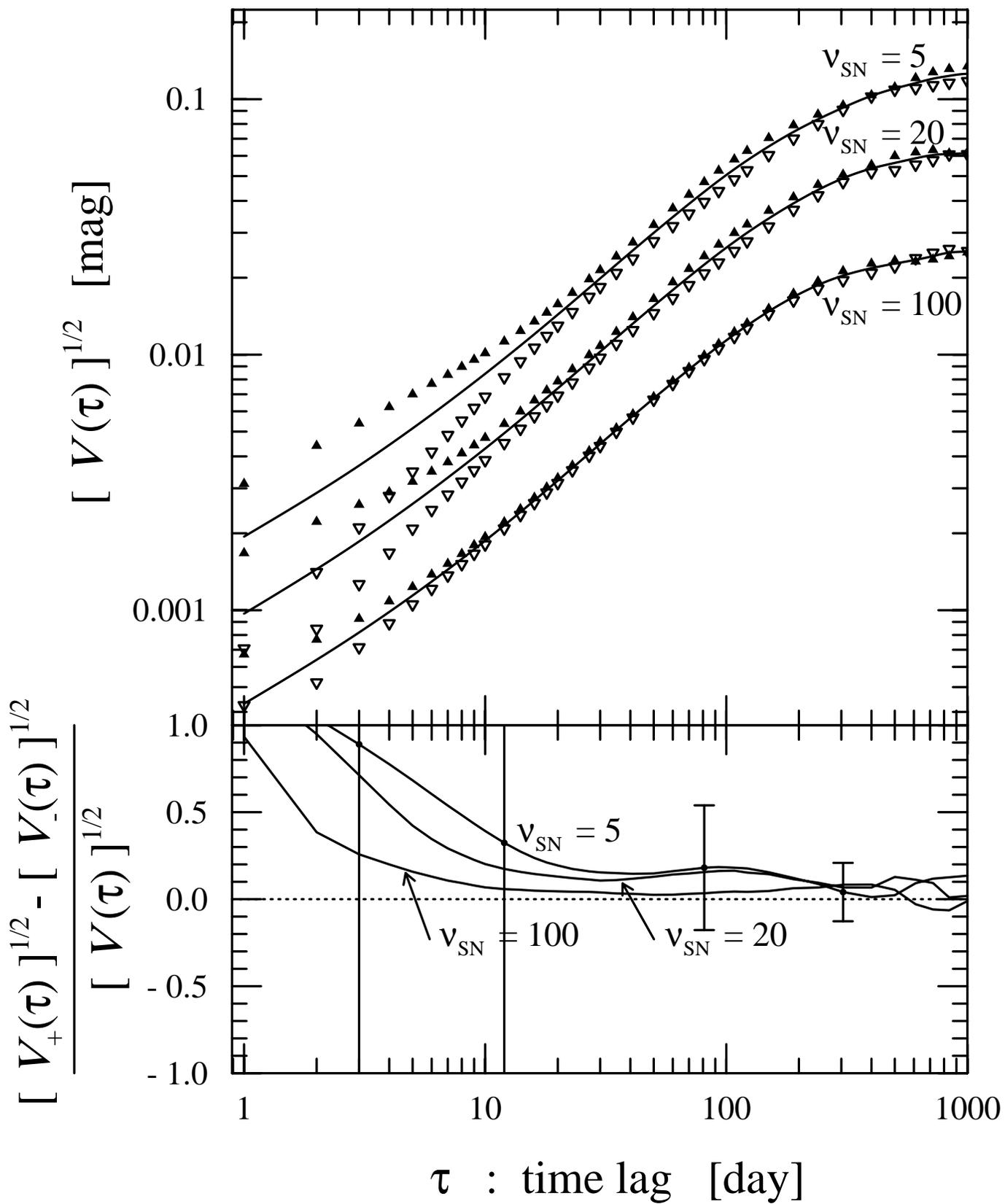
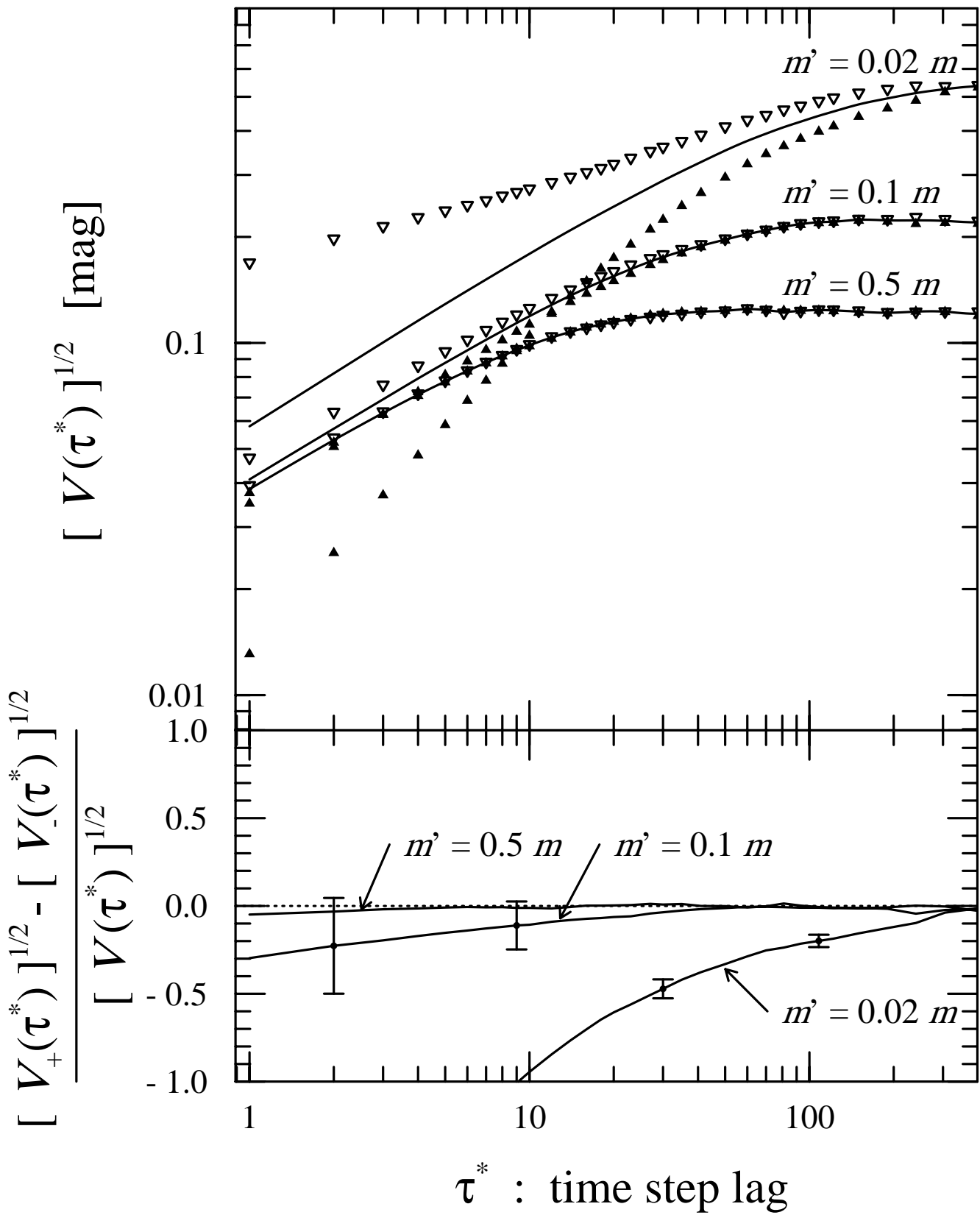
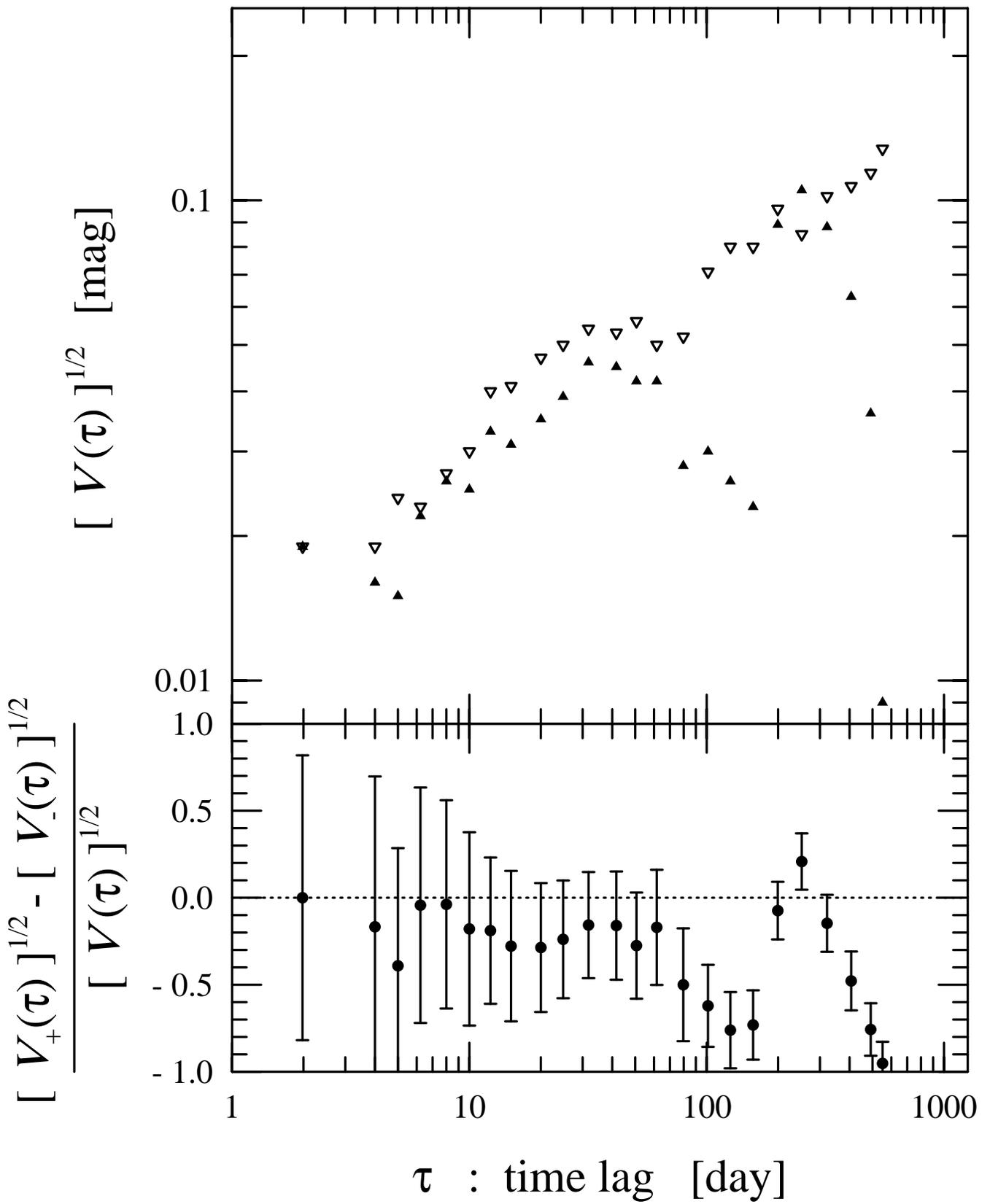
\end{document}